\documentstyle[eqsecnum,prd,aps,epsf]{revtex}

\newcommand{\gsim}
   {\mathrel{\raise.3ex\hbox{$>$\kern-.75em\lower1ex\hbox{$\sim$}}}}
\newcommand{\lsim}
   {\mathrel{\raise.3ex\hbox{$<$\kern-.75em\lower1ex\hbox{$\sim$}}}}

\draft
\begin{document}

\thispagestyle{empty}
{\baselineskip0pt
\leftline{\large\baselineskip16pt\sl\vbox to0pt{\hbox{\it Department of Physics}
               \hbox{\it Kyoto University}\vss}}
\rightline{\large\baselineskip16pt\rm\vbox to20pt{\hbox{KUNS 1439}
           \hbox{April 12, 1997}
\vss}}%
}
\vskip1cm
\begin{center}{\large \bf
Stability Analysis of Spherically Symmetric Star in 
Scalar-Tensor Theories of Gravity
}
\end{center}
\vskip1cm
\begin{center}
 {\large 
Tomohiro Harada
\footnote{ Electronic address: harada@tap.scphys.kyoto-u.ac.jp} } \\
{\em Department of Physics,~Kyoto University,} 
{\em Kyoto 606-01,~Japan}\\
\end{center}

\begin{abstract}
A stability analysis of a spherically symmetric star
in scalar-tensor theories of gravity is given in 
terms of the frequencies of quasi-normal modes.
The scalar-tensor theories have a scalar field which
is related to gravitation.
There is an arbitrary function, the so-called coupling function,
which determines the strength of the coupling between the 
gravitational scalar field and matter.
Instability is induced by the scalar field
for some
ranges of the value of the first derivative of the coupling
function.
This instability leads to significant discrepancies
with the results of binary-pulsar-timing experiments
and hence, by the stability analysis, we can
exclude the ranges of the first derivative of the 
coupling function in which the instability sets in.
In this article, the constraint on the first derivative of the
coupling function from the stability of relativistic stars
is found.
Analysis in terms of the quasi-normal mode frequencies
accounts for the parameter dependence of 
the wave form of the 
scalar gravitational waves emitted 
from the Oppenheimer-Snyder
collapse.
The spontaneous scalarization
is also discussed.

\end{abstract}


\section{INTRODUCTION}
General relativity is not the 
only viable theory of gravity.
Scalar-tensor theories of gravity are among the
alternative theories.
The scalar-tensor theories contain a 
gravitational scalar field
and an arbitrary function that
determines the strength of the coupling between 
the scalar field and matter.
With some choice of the coupling function,
the scalar-tensor theories can pass all
the present experimental tests of gravitation.
From a theoretical point of view,
it has been pointed out that
the scalar-tensor theories arise naturally as the
low-energy limit of string 
theory~\cite{callan,damour-polyakov}.
Until now almost 
all tests of high 
precision have been performed
only in a weak field. 
It has been pointed out, however, 
by Damour and 
Esposito-Far\`ese~\cite{damour-esposito-farese2,damour-esposito-farese3}
that there are 
considerable deviations 
of predictions in a strong field 
given by some viable scalar-tensor theories from those 
given by general relativity.

In the scalar-tensor theories of gravity,
as in the case of general relativity,
generation of gravitational waves is a typical phenomenon
in a strong field and hence
it is expected~\cite{will,shibata} that the projects
(LIGO~\cite{ligo}, 
VIRGO~\cite{virgo}, GEO~\cite{geo} and
TAMA~\cite{tama}) of gravitational wave observations
by laser interferometers
will test the viable 
gravitational theories with high accuracy.
Unlike general relativity, there is a
scalar mode in gravitational waves in the
scalar-tensor theories.
Since this scalar gravitational wave can be detected 
by laser interferometers~\cite{shibata} or 
resonant mass antennas and 
can be separated from tensor modes,
it has crucial importance with respect to the observational
tests of the theories of gravity.
The scalar gravitational waves from the
Oppenheimer-Snyder collapse,
i.e., collapse of a spherically symmetric and
homogeneous dust ball, have been calculated 
numerically as reported in Ref.~\cite{harada},
where behavior of the scalar field
within the dust, the observed wave form 
of the scalar gravitational waves,
and its dependence on the value of the 
first derivative of the coupling function
are examined.
From the observed wave form of the
scalar gravitational waves we can 
obtain information concerning the
first derivative of the coupling function.
The dependence of the observed 
wave form on the first derivative of the
coupling function can be explained by stability analysis
of the quasi-normal modes
since the behavior of the scalar field under the 
outgoing wave condition
before the formation of the horizon 
is characterized by the quasi-normal modes
that satisfy the regularity condition at the center.

The stability analysis is motivated
by the gravitational test through pulsar-timing experiments.
By measuring the change of the period of pulsation
we can obtain information regarding 
scalar ``form factors'' of the pulsar, and
those form factors are related to the
variation of the moment of inertia 
under the influence of the companion
star~\cite{damour-esposito-farese3}.
It was shown that in some class of the scalar-tensor
theories there is a solution with
non-perturbative strong-field 
effects
and it has form factors that are significantly
different from those in general 
relativity~\cite{damour-esposito-farese2,damour-esposito-farese3}.
In order to obtain the correct gravitational theory,
it is important to investigate the internal structure
of highly relativistic stars and
check the consistency with the results observational tests.
The stability analysis reveals whether the deviation of the
stable equilibrium solution of the relativistic star
from that in general relativity
remains small or not.
If, with some choice of the coupling function,
the stable equilibrium solution suffers from 
non-perturbative effects by
the coupling between the scalar field and matter
and has significantly different scalar form factors
from those in general relativity,
the results of the pulsar-timing experiments
will exclude that choice of the coupling function.

The condition for the stability of 
relativistic stars of a perfect fluid 
in the scalar-tensor theories for radial and time-symmetric
perturbations was examined by Bruckman 
and Vel\'azquez~\cite{bruckman-velazquez}.
They found that the stability depends on the
choice of the coupling function.

This paper is organized as follows.
In Sec. \ref{sec:basiceq} we review the scalar-tensor theories 
of gravity and
derive the basic equation.
In Sec. \ref{sec:stabilityanalysis} we
introduce the quasi-normal modes of the scalar field
and discuss some useful conditions concerning them.
In Sec. \ref{sec:models} we introduce stellar models.
In Sec. \ref{sec:numresults} we give 
numerical results on stability versus instability for
these stellar models.
In Sec. \ref{sec:summary} we summarize the results
and discuss their implications. 
We use the units in which $c=1$.
We follow the MTW~\cite{mtw} sign conventions for the
metric tensor, Riemann tensor and Einstein tensor.

\section{BASIC EQUATION}
\label{sec:basiceq}

\subsection{Scalar-Tensor Theories of Gravity}We consider a class of scalar-tensor theories of gravity
in which gravitation is mediated by a
long-range scalar field in addition to space-time curvature.
The action in the Einstein frame formulation 
is given by~\cite{damour-esposito-farese1}
\begin{equation}
    I=\frac{1}{16\pi G_*}\int\sqrt{-g_*}\left( R_* -
    2g_*^{\mu\nu}\varphi_{,\mu}\varphi_{,\nu}
    \right)d^4x
    +I_m[\Psi_m,A^2(\varphi)g_{*\mu\nu}], 
\end{equation}
where the metric tensor $g_{*\mu\nu}$ in the ``unphysical''
Einstein frame is related to the metric tensor $g_{\mu\nu}$
in the ``physical'' Brans-Dicke frame by the conformal
transformation as  
\begin{equation}
      g_{\mu\nu}=A^2(\varphi)g_{*\mu\nu}.
      \label{eq:conformaltransformation}
\end{equation}
Here, $G_*$ is the ``bare'' gravitational constant and
$\varphi$ is the gravitational scalar field.
$R_*$ is the scalar curvature of $g_{*\mu\nu}$,
and $\Psi_m$ represents matter fields collectively.
The arbitrary function $A(\varphi)$ is related
to a coupling function which will be introduced below.
The Brans-Dicke frame metric $g_{\mu\nu}$ is
universally coupled to matter in the model 
we consider here,
and hence the Einstein equivalence principle is valid.

The field equations are given by
\begin{eqnarray}
      & & G_{*\mu\nu}=8\pi G_* T_{*\mu\nu}
      +2\left(\varphi_{,\mu}\varphi_{,\nu}-\frac{1}{2}g_{*\mu\nu}
      g_*^{\alpha\beta}\varphi_{,\alpha}\varphi_{,\beta}\right), 
    \label{eq:fieldeq1} \\
      & & \Box_* \varphi = - 4\pi G_* \alpha(\varphi) T_*,
      \label{eq:fieldeq2}
\end{eqnarray}
and the equations of motion are given by
\begin{equation}
  \label{eq:eom}
  \nabla_{*\nu}T_{*\mu}^{\nu}=\alpha(\varphi)T_*\nabla_{*\mu}\varphi,
\end{equation}
where
\begin{eqnarray}
      T_*^{\mu\nu} &\equiv& \frac{2}{\sqrt{-g_*}}
      \frac{\delta I_m[\Psi_m,A^2(\varphi)g_{*\mu\nu}]}
      {\delta g_{*\mu\nu}}
      = A^6(\varphi) T^{\mu\nu}, \\
      T_* &\equiv& T_{*\mu}^{~~\mu} 
      \equiv T^{\mu\nu}_* g_{*\mu\nu},\\
      \alpha(\varphi) &\equiv& \frac{d\ln A(\varphi)}
      {d\varphi}.
\end{eqnarray}
In the above equations,
$G_{*\mu\nu}$, $\nabla_{*\nu}$ and $\Box_*$ are
the Einstein tensor, covariant derivative 
and d'Alembertian of $g_{*\mu\nu}$
respectively,
and $T^{\mu\nu}$ is the stress-energy tensor of matter in the
Brans-Dicke frame.

The function $\alpha(\varphi)$ is a coupling
function between the scalar field and trace of the 
stress-energy tensor of matter,
as seen in Eq.~(\ref{eq:fieldeq2}).
If $\alpha(\varphi)$ is constant, the theory reduces to
the Brans-Dicke theory~\cite{brans-dicke}.
If $\alpha(\varphi)=0$ , the theory reduces to
general relativity.
Hereafter we consider an asymptotically flat space-time
and we assume $\varphi \to \varphi_0$ at spatial infinity.
We use units in which $G_*=1$, and 
we fix the freedom of the constant conformal transformation 
by requiring $A(\varphi_0)=1$.

Here we briefly summarize the
present constraints on the coupling function $\alpha(\varphi)$ 
through the results of solar-system test experiments.
The solar system is a laboratory in a weak gravitational field, 
and hence
the parametrized-post-Newtonian (PPN) formalism~\cite{will1}
works well.
In the scalar-tensor theories introduced above,
the so-called Eddington parameters among the PPN parameters 
are expressed by the coupling function $\alpha(\varphi)$ 
as~\cite{damour-esposito-farese1}
\begin{eqnarray}
      1-\gamma_{Edd} &=& \frac{2\alpha_0^2}{1+\alpha_0^2}, \\
      \beta_{Edd}-1 &=& \frac{\beta_0\alpha_0^2}{2(1+\alpha_0^2)^2},
\end{eqnarray}
where
\begin{eqnarray}
  \label{eq:defofalpha0}
      \alpha_0 &\equiv& \alpha(\varphi_0), \\
      \label{eq:defofbeta0}
      \beta_0 &\equiv& \frac{d\alpha}
      {d\varphi}(\varphi_0),
\end{eqnarray}
and the values of the other PPN parameters 
that enter in the first post-Newtonian approximation are
identical to those for general relativity~\cite{will1}.
The results of the experiments of the
light deflection constrain $\gamma_{Edd}$
as~\cite{lebach}
\begin{equation}
      \gamma_{Edd}= 0.9996\pm0.0017,
\end{equation}
and this constraint reduces to 
\begin{equation}
  \label{eq:constraint1}
      \alpha_0^2<0.001.
\end{equation}
The results of the lunar laser-ranging experiments
constrain $\beta_{Edd}$ as~\cite{damour}
\begin{equation}
  \beta_{Edd}=0.9998\pm0.0006,
\end{equation}
and this constant reduces to
\begin{equation}
  \label{eq:constraint2}
  \beta_0\alpha_0^2=-0.0004\pm0.0012.
\end{equation}

\subsection{Perturbation Equations}
From this point we restrict our attention to the scalar-tensor 
theories in which
\begin{equation}
  \alpha_0 = 0.
\end{equation}
This choice is completely consistent with the experimental constraints
Eqs. (\ref{eq:constraint1}) and (\ref{eq:constraint2}).
The scalar-tensor theories in which $\alpha_0=0$ 
pass all weak-field tests of solar-system experiments 
that have been carried out until now,
for an arbitrary value
of $\beta_0$.
Further by this assumption we can seek
the pure effect of the first derivative $\beta_0$ 
of the coupling function $\alpha(\varphi)$ on 
the stability of a star. 

For $\alpha_0=0$, the field equations (\ref{eq:fieldeq1})
and (\ref{eq:fieldeq2})
allow the solution 
\begin{eqnarray}
  g_{*\mu\nu}&=&g_{\mu\nu}^{(E)}, \\
  T_{*\mu\nu}&=&T_{\mu\nu}^{(E)}, \\
  \varphi&=&\varphi_0 ,
\end{eqnarray}
where the set of $g_{\mu\nu}^{(E)}$ and $T_{\mu\nu}^{(E)}$ is
a solution of the Einstein equation, i.e.,
\begin{equation}
  G_{\mu\nu}^{(E)}=8\pi T_{\mu\nu}^{(E)},
\end{equation}
and therefore
\begin{equation}
  \label{eq:unperturbedeom}
  \nabla^{(E)}_{\nu}T_{\mu}^{(E)\nu}=0.
\end{equation}
Here, $\nabla^{(E)}_{\nu}$ and $G_{\mu\nu}^{(E)}$
are the covariant derivative and Einstein tensor
of $g_{\mu\nu}^{(E)}$, respectively.
In this solution the Einstein frame expression agrees
with the Brans-Dicke frame expression:
\begin{eqnarray}
  g_{\mu\nu} &=& g_{*\mu\nu}, \nonumber \\
  T_{\mu\nu} &=& T_{*\mu\nu}. \nonumber
\end{eqnarray}

If we consider linear perturbations
of this solution, the coupled
differential equations for the 
perturbations
decouple to an equation for the scalar field perturbation
and equations for the metric and matter perturbations.
This is because, for $\alpha_0=0$,
the perturbation of the conformal 
factor $A^{2}(\varphi)$ defined in 
Eq.~(\ref{eq:conformaltransformation}) 
by the scalar field perturbation from
the constant value $\varphi_0$
is a second-order infinitesimal.
For the same reason, the perturbations 
$\delta g_{*\mu\nu}$ and $\delta T_{*\mu\nu}$ 
of $g_{*\mu\nu}$ and
$T_{*\mu\nu}$ in the Einstein frame 
are the same as  
$\delta g_{\mu\nu}$ and $\delta T_{\mu\nu}$ 
of $g_{\mu\nu}$ and
$T_{\mu\nu}$ in the Brans-Dicke frame
up to linear order:  
\begin{eqnarray}
  g_{*\mu\nu} &=& g_{\mu\nu}^{(E)}+\delta g_{\mu\nu}+\mbox{higher
order terms}, \\
  T_{*\mu\nu} &=& T_{\mu\nu}^{(E)}+\delta T_{\mu\nu}+\mbox{higher
order terms}, \\
  \varphi &=& \varphi_0 + \delta\varphi.
\end{eqnarray}
From Eqs. (\ref{eq:fieldeq1}), (\ref{eq:fieldeq2})
and (\ref{eq:eom}), 
we obtain 
the field equations for the linear perturbations as
\begin{eqnarray}
  \delta G_{\mu\nu} &=& 8\pi \delta T_{\mu\nu}, 
  \label{eq:eqmetric} \\
  \Box^{(E)}\delta\varphi &=& -4\pi \beta_0 T^{(E)}\delta\varphi,
  \label{eq:waveeq}
\end{eqnarray}
and the equations of motion as
\begin{equation}
  \label{eq:perturbedeom}
  \delta{(\nabla_{\nu}T_{\mu}^{\nu})}=0,
\end{equation}
where $\Box^{(E)}$ is the d'Alembertian of $g_{\mu\nu}^{(E)}$.
Equations (\ref{eq:eqmetric}) for the metric and 
matter perturbations 
are identical to those for general relativity.
Since we are now interested in 
instability induced by the gravitational scalar field
we do not consider these equations here.
Thus it is sufficient for our purposes to
consider only Eq.~(\ref{eq:waveeq}) for the
scalar field perturbation.  
In what follows we omit the symbol $(E)$ 
to simplify the notation.

We assume that the unperturbed solution is
static and spherically symmetric.
Then, without loss of generality, the metric
is written in the simple form
\begin{equation}
  \label{eq:staticspherical}
    ds^2 = -e^{2\Phi(r)}dt^2 + e^{2\Psi(r)}dr^2 + r^2 (d\theta^2 +
    \sin^2 \theta d\phi^2). 
\end{equation}
In this space-time, Eq.~(\ref{eq:waveeq}) becomes
\begin{eqnarray}
  & &-e^{-2\Phi}\frac{\partial^2 \delta\varphi}{\partial t^2}
  +\frac{e^{-\Phi-\Psi}}{r^2}\frac{\partial}{\partial r}
  \left( e^{\Phi-\Psi}r^2\frac{\partial \delta\varphi}{\partial r}
  \right) \nonumber \\
  & &+\frac{1}{r^2}\left[\frac{1}{\sin\theta}\frac{\partial}
  {\partial\theta}\left(\sin\theta\frac{\partial\delta\varphi}
  {\partial\theta}\right)+\frac{1}{\sin^2\theta}
  \frac{\partial^2\delta\varphi}{\partial\phi^2}\right] 
  +4\pi\beta_0 T \delta\varphi=0.
  \label{eq:explicitwaveeq}
\end{eqnarray}

We concentrate on perturbations 
with sinusoidal time
dependence, and define $\psi_{\omega l m}(r)$ as
\begin{equation}
  \label{eq:sinusoid}
  \delta\varphi=e^{i\omega t}\sum_{l,m}
  \frac{\psi_{\omega l m}(r)}{r} 
  Y^{l}_{m}(\theta,\phi),
\end{equation}
where we have followed the sign convention 
of Ref.~\cite{chandrasekhar}.
Hereafter we abbreviate $\psi_{\omega l m}(r)$
as $\psi(r)$.
Then from Eq.~(\ref{eq:explicitwaveeq}) we obtain
a Schr\"odinger-like equation as
\begin{equation}
  \frac{d^2 \psi}{d r_*^2}+[\omega^2-V(r_*)]\psi=0 ,
  \label{eq:schroedinger}
\end{equation}
where we have defined the ``generalized tortoise coordinate'' $r_*$ as
\begin{equation}
  dr_* \equiv e^{\Psi-\Phi}dr.
\end{equation}
The effective potential, $V(r_*)$, is defined as
\begin{equation}
  V(r_*) \equiv \frac{\Phi^{\prime}-\Psi^{\prime}}{r}e^{2(\Phi-\Psi)}
         +\frac{l(l+1)}{r^2}e^{2\Phi}
         -4\pi \beta_0 T e^{2\Phi},
         \label{eq:potential}
\end{equation}
where the prime denotes $d/dr$ and $\beta_0$ has been
defined in Eq.~(\ref{eq:defofbeta0}).
On the right-hand side of 
Eq.~(\ref{eq:potential}), the first term is a ``curvature
potential'', the second is a ``centrifugal potential'',
and the third
comes from the fact that the coupling function $\alpha(\varphi)$
depends on the scalar field $\varphi$,
and hence $\alpha(\varphi)$ is not constant when the
perturbation of the scalar field $\varphi$ is present.
  
\section{STABILITY ANALYSIS}
\label{sec:stabilityanalysis}

\subsection{the Quasi-Normal Modes}
For any complex number $\omega$,
a general solution of Eq.~(\ref{eq:schroedinger})
consists of a linear combination of two independent 
solutions as
\begin{equation}
  \label{eq:linearcomb}
  \psi(r_*)=C^{+}(\omega)\psi^{+}(r_*)+C^{-}(\omega)\psi^{-}(r_*),
\end{equation}
where $\psi^{+}$ and $\psi^{-}$ are
purely ingoing and outgoing waves. That is,
\begin{eqnarray}
\psi^{+} &\longrightarrow& e^{i\omega r_*}, \\
\psi^{-} &\longrightarrow& e^{-i\omega r_*},
\end{eqnarray}
in the limit $r_*\to \infty$.
The coefficients $C^{+}(\omega)$ and $C^{-}(\omega)$
are determined by
\begin{eqnarray}
  \label{eq:coefficientsab}
C^{+}(\omega) &=& \frac{W(\psi,\psi^{-})}{W(\psi^{+},\psi^{-})},\\
C^{-}(\omega) &=& \frac{W(\psi,\psi^{+})}{W(\psi^{-},\psi^{+})},
\end{eqnarray}
where $W(f,g)$ is the Wronskian defined as
\begin{equation}
W(f,g)\equiv f\frac{dg}{dr_*}-\frac{df}{dr_*}g.
\end{equation}

For a general complex number $\omega$ there may be
no solution that satisfies the 
regularity condition ($\psi=0$) at the 
origin and outgoing wave boundary condition
at infinity ($C^{+}(\omega)=0$) simultaneously.
For some special, discrete values of 
the frequency $\omega$
there exists a purely outgoing wave solution
that satisfies the regularity condition.
We have special interest in this solution
as an eigenmode of the scalar field.
We call the solution a quasi-normal mode
and this special complex number $\omega$ 
a quasi-normal mode frequency.
The physical meaning of the quasi-normal mode
of the scalar field is that when the scalar field
is perturbed by some effect other than the incidence of
scalar waves, the scalar field oscillates 
with radiation of
scalar waves, and its characteristic oscillation
modes are described by the quasi-normal modes.

\subsection{Stability of the Quasi-Normal Modes}
With the sign convention in 
Eq.~(\ref{eq:sinusoid}), if the imaginary part of 
the quasi-normal mode frequency $\omega$
is positive, the quasi-normal oscillation is damped,
and its damping rate is given by this imaginary part.
If the imaginary part of $\omega$ is negative, however,
the quasi-normal mode is unstable, and the amplitude of 
the perturbation
grows exponentially in the growth rate given by the 
absolute value of the imaginary part. 
The unstable quasi-normal mode develops from
a regular initial set of data, because the
amplitude of the perturbation decreases exponentially
to zero in the limit $r_*\to \infty$.
Whether the quasi-normal mode is stable or not
is thus determined by whether its complex 
frequency is in the upper half plane
or in the lower half plane.

As for the stability of the quasi-normal modes of 
Eq.~(\ref{eq:schroedinger}),
the following conditions, similar to those for the 
quasi-normal modes of spherical
nonrotating stars in general 
relativity~\cite{thorne}, hold:
(i) An unstable quasi-normal mode of any $l$
has a purely imaginary frequency, i.e.,
an unstable mode grows exponentially in time
without oscillation.
(ii) If the effective potential $V$ is non-negative everywhere,
there is no unstable quasi-normal mode.
(iii) Consider a sequence of effective potentials
smoothly parametrized by a real variable $\lambda$.
Then there are discrete quasi-normal mode frequencies, and
each quasi-normal mode frequency, 
denoted by $\omega_n(\lambda)$, 
moves smoothly along a trajectory in the complex plane
as $\lambda$ varies smoothly.
(iv) A quasi-normal mode frequency is bounded.
(v) 
Consider a sequence of effective potentials
smoothly parametrized by a real variable $\lambda$.
If there is a critical value $\lambda_n^{crit}$
for any quasi-normal mode frequency $\omega_n$
such that $\mbox{Im}[\omega_n(\lambda)]>0$ for
$\lambda>\lambda_n^{crit}$ and $\mbox{Im}[\omega_n(\lambda)]<0$
for $\lambda<\lambda_n^{crit}$,
then $\omega_n(\lambda_n^{crit})=0$.
In other words, 
when the stability of 
a quasi-normal mode changes 
as $\lambda$ varies smoothly,
its frequency is zero. 

Condition (i) is the direct result of the fact that 
the finiteness of the norm of an
unstable quasi-normal mode and hermitianity of the 
derivative operator 
\begin{equation}
-\frac{d^2}{dr_*^2}+V,
\end{equation}
guarantee that the eigenvalue $\omega^2$
is real for the unstable quasi-normal mode.
Condition (ii) is equivalent to the absence of bound states
in the non-negative potential in quantum mechanics.
The proof of condition (iii) is almost the same as that
for the quasi-normal mode frequencies of
gravitational waves in general 
relativity~\cite{detweiler-ipser}.
The sketch of the proof is as follows.
First we take $u\equiv r^{-l-1}\psi$ in place of $\psi$ (cf. 
Appendix \ref{sec:regularity}).
The regularity condition requires $du/dr=0$ at the origin,
and we fix the normalization by requiring $u=1$ at the origin.
The solutions $u$, and therefore 
$\psi$ depend smoothly on $\lambda$
because each term in the differential equation 
(\ref{eq:schroedinger}) 
depends smoothly on $\lambda$.
So $C^{+}(\omega;\lambda)$, which is defined by $C^{+}(\omega)$ 
for $\lambda$, is a smooth function of $\lambda$. 
The solutions $u$ and therefore $\psi$
are also analytic functions of $\omega$ because
the analyticity of each term for $\omega$ in the 
differential equation (\ref{eq:schroedinger}) 
guarantees the Cauchy-Riemann relations for its solutions.
Therefore $C^{+}(\omega;\lambda)$ is an analytic function of $\omega$.
From the property of an analytic function,
any zero of $C^{+}(\omega;\lambda)$ for any given $\lambda$,
denoted by $\omega_n(\lambda)$, 
is isolated,
and the number of the zeroes is, at most, countably infinite.
If we were to assume the discontinuity 
of each quasi-normal mode
frequency $\omega_n(\lambda)$ with respect to $\lambda$,  
a contradiction would result with the smoothness
of $C^{+}(\omega;\lambda)$ with respect to 
$\lambda$ through the maximum principle. 
Therefore we conclude that each quasi-normal mode 
frequency $\omega_n(\lambda)$ 
varies smoothly with respect to $\lambda$. 
Condition (iv) can be easily shown by observing 
that in the limit $|\omega|\to\infty$,
the solution regular at the origin is not
a purely outgoing wave.
Condition (v) is the combined result of 
conditions (i), (iii) and (iv).

We define a critical value $\beta_0^{crit}$ as a
value of $\beta_0$ such that for $\beta_0>(<)\beta_0^{crit}$
there is no unstable mode and for $\beta_0<(>)\beta_0^{crit}$
there is an unstable mode.
The corollary of condition 
(ii) is that if $T=T^{\mu}_{\mu}<0$
in the stellar interior, which is considered to
be satisfied in most of the physical situations,
then there is a critical value
$\beta_0^{crit}$, and for any $\beta_0$ larger than this value
there is no unstable quasi-normal mode.
The corollary of condition (iii) is that 
each quasi-normal mode frequency moves smoothly
in the complex plane
as we change $\beta_0$ smoothly. 
The corollary of condition (v) is that
there is a zero-frequency quasi-normal mode
at $\beta_0=\beta_0^{crit}$. 
\section{MODELS}
\label{sec:models}
To obtain further understanding
of the instability induced by the scalar field
we consider stellar models that describe
relativistic stars.
First we take as a stellar model the equilibrium
solution of an incompressible fluid
(see Box 23.2 of MTW~\cite{mtw}).
In this case the matter is a perfect fluid:
\begin{equation}
  \label{eq:matter}
    T_{\mu\nu}=\rho u_{\mu}u_{\nu}+P(g_{\mu\nu}+u_{\mu}u_{\nu}). 
\end{equation}
The trace of the stress-energy tensor is given by
\begin{equation}
  \label{eq:trace}
  T =  T^{\mu}_{~\mu}=-\rho+3P.
\end{equation}
The energy density and pressure are given by
\begin{eqnarray}
  \label{eq:density}
    \rho(r) &=& \cases{
              \rho_0 & ($0\leq r < R$) \cr
              0      & ($R\leq r$)     \cr
              },\\ 
  \label{eq:pressure}
    P(r) &=& \cases{
              \rho_0\left[\frac{\left(1-\frac{2Mr^2}{R^3}\right)^{1/2}
              -\left(1-\frac{2M}{R}\right)^{1/2}}
              {3\left(1-\frac{2M}{R}\right)^{1/2}-
              \left(1-\frac{2Mr^2}{R^3}\right)^{1/2}}
              \right] & ($0\leq r < R$) \cr
              0      & ($R\leq r$)     \cr
              },
\end{eqnarray}
where $M$ and $R$ are the gravitational mass and radius of the star,
\begin{equation}
  \label{eq:mass}
    M = \frac{4}{3}\pi\rho_0 R^3.
\end{equation}
The metric is given in the form of 
Eq.~(\ref{eq:staticspherical}), where
\begin{eqnarray}
    e^{2\Phi(r)} &=& \cases{\frac{1}{4}
                 \left[3\left(1-\frac{2M}{R}\right)
                 ^{1/2}-\left(1-\frac{2Mr^2}
                 {R^3}\right)^{1/2}\right]^2 
                 & ($0\leq r < R$) \cr
                 1-\frac{2M}{r}
                 & ($R\leq r$)     \cr 
              }    ,\\
            \label{eq:defofmass}
    e^{2\Psi(r)} &=& \left[1-\frac{2m(r)}{r}\right]^{-1} ,\\
    m(r) &=& \cases{ 
           \frac{4}{3}\pi\rho_0 r^3
             & ($0\leq r < R$) \cr
           M & ($R\leq r$)     \cr
           }.  
\end{eqnarray}
There exists a smallest radius of a star of homogeneous density.
From this, we have the condition
\begin{equation}
  \label{eq:lowestradius}
  \frac{R}{M} > \frac{9}{4} .
\end{equation}
We can easily see that 
the negativity of $T$ in the stellar 
interior holds if and only if
\begin{equation}
  \label{eq:negativet}
  \frac{R}{M} > \frac{18}{5}, 
\end{equation}
and 
the dominant energy condition holds if and only if
\begin{equation}
  \label{eq:dec}
  \frac{R}{M} > \frac{8}{3}. 
\end{equation}

Next we introduce a polytropic stellar model.
The metric for a static and spherically symmetric space-time
is written in the form of Eq.~(\ref{eq:staticspherical}).
The stress-energy tensor of a perfect fluid is given 
by Eq.~(\ref{eq:matter}).
We integrate the Tolman-Oppenheimer-Volkoff equation
\begin{equation}
  \label{eq:tov}
  \frac{dP}{dr}=-(P+\rho)\frac{m+4\pi r^3 P}{r[r-2m]},
\end{equation}
where $m(r)$ is defined by Eq.~(\ref{eq:defofmass})
and determined by
\begin{equation}
  \label{eq:qlmass}
  m(r)=4\pi\int^r_0\rho(r^{\prime})r^{\prime 2}dr^{\prime}.
\end{equation}
The function $\Phi(r)$ is determined by integrating the differential equation
\begin{equation}
  \label{eq:phi}
  \frac{d\Phi}{dr}=-\frac{1}{P+\rho}\frac{dP}{dr},
\end{equation}
and the requirement of matching to the Schwarzschild space-time at the
stellar surface.
We consider the following polytropic equation of state:
\begin{eqnarray}
  \rho &=& nm_b+\frac{Kn_0m_b}{\Gamma-1}
  \left(\frac{n}{n_0}\right)^{\Gamma}, \\
  P &=& Kn_0m_b\left(\frac{n}{n_0}\right)^{\Gamma}, \\
  m_b &=& 1.66\times10^{-24}\mbox{g}, \\
  n_0 &=& 0.1\mbox{fm}^{-3}.
\end{eqnarray}
We then take the parameters $\Gamma=2.34$ and 
$K=0.0195$~\cite{damour-esposito-farese3}
to fit the more realistic equation of 
state for high-density nuclear 
matter~\cite{diaz-alonzo-ibanez-cabanell}.

\begin{figure}[htb]
  \epsfysize 6cm \epsfxsize 8cm
  \centerline{\epsfbox{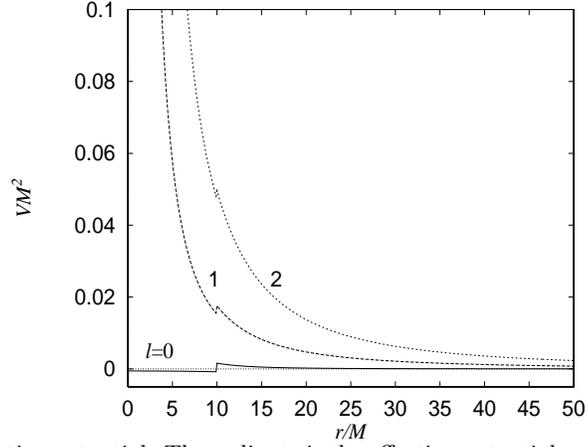}}
\caption{$l$-dependence on the effective potential.
The ordinate is the effective potential and
the abscissa is the areal coordinate.
The stellar radius $R$ is fixed to $10M$.
The theoretical parameter $\beta_0$ is fixed to 0.}
\end{figure}

\begin{figure}[htb]
  \epsfysize 6cm \epsfxsize 8cm
  \centerline{\epsfbox{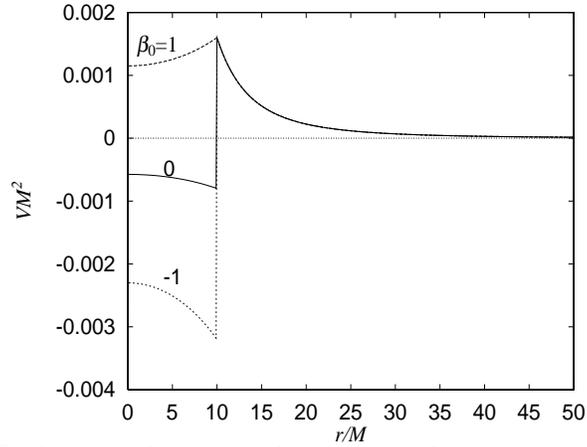}}
\caption{$\beta_0$-dependence on the effective
potential.
The ordinate and abscissa are
the same as in Fig. 1.
The stellar radius $R$ is fixed to $10M$.
The angular momentum $l$
is fixed to 0.}
\end{figure}
\begin{figure}[htb]
  \epsfysize 6cm \epsfxsize 8cm
  \centerline{\epsfbox{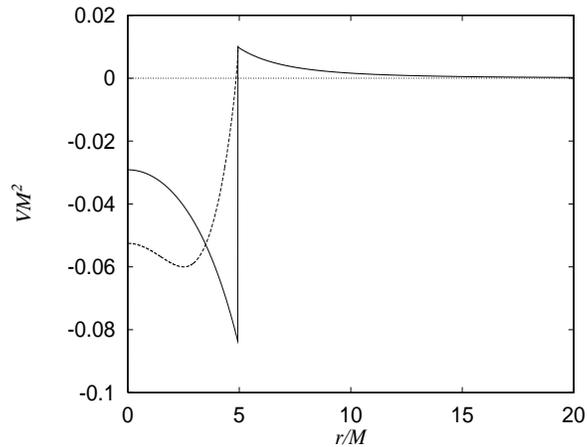}}
  \caption{Comparison between the effective potentials of an $l=0$ mode
for $\beta_0=-5$ of the
incompressible fluid model and polytropic model.
The ordinate and abscissa are the same as in Fig. 1.
The solid and dashed lines denote the
incompressible fluid model
and polytropic model respectively.
In the polytropic model, the total
central density is $1.16\times10^{15}\mbox{g/cm}^3$,
and the stellar mass is $1.74M_{\odot}$.
The stellar radius is $4.94M$ in both models.
The details of the polytropic stellar model are given
in the Sec. \protect\ref{sec:models} and Appendix 
\protect\ref{sec:poly}.}
\end{figure}

The effective potentials of the stellar model
of an incompressible fluid
are shown in Figs. 1, 2 and 3.
Figure 1 shows the $l$-dependence of the shape of the 
effective potential of the incompressible
fluid model for $R=10M$.
The parameter $\beta_0$ is fixed to 0.
The centrifugal potential raises significantly the 
total effective potential
for $l\ge 1$ modes.
Therefore, from condition (i), an $l=0$ 
(spherically symmetric) mode is 
more likely to be unstable than
any $l\ge 1$ mode, and therefore
we concentrate our attention to the $l=0$ mode.  
Figure 2 shows the $\beta_0$-dependence of the shape of
the effective potential 
of the incompressible fluid model for $l=0$ and $R=10M$.
The term which comes from 
the coupling between the scalar field and matter
depends on $\beta_0$.
As $\beta_0$ becomes smaller, the quasi-normal modes
become more likely to be unstable
because the potential well
becomes deeper.
In particular, from condition (i) and Fig. 2, 
no unstable quasi-normal mode exists
in the theory in which $\beta_0=1$ for $R=10M$
in this model.
Figure 3 shows the effective potential
of the stellar model of an incompressible fluid 
and that of the polytropic stellar model
of an $l=0$ mode for $\beta_0=-5$. 
In Fig. 3 we fix the central total energy density to be 
$1.16\times 10^{15}\mbox{g/cm}^{3}$ in the 
polytropic stellar model 
and obtain the stellar radius $12.7\mbox{km}$, 
mass $1.74M_{\odot}$ and radius-to-mass ratio $4.94$.
We plot, for comparison, 
the effective potential of the 
model of an incompressible fluid 
whose radius-to-mass ratio 
is also fixed to 4.94.
The details of the 
numerical calculation of
the polytropic stellar model
are described in 
Appendix \ref{sec:poly}.

\section{NUMERICAL RESULTS}
\label{sec:numresults}
To search for $\beta_0^{crit}$ we integrate
the differential equation (\ref{eq:schroedinger}) for $\omega=0$ 
from the origin
under the regularity condition.
The regularity condition is given by the Taylor
expansion of the solution around the origin.
The expansion 
of the regular solution around the origin 
is described in Appendix \ref{sec:regularity}.
After we integrate Eq.~(\ref{eq:schroedinger})  
to the stellar surface
we calculate the Wronskian
with an exterior solution that is
regular at $r=\infty$. 
As shown in Appendix \ref{sec:exterior},
we know that 
the exterior solution for $\omega=0$ that is regular at
$r=\infty$ is given by the 
hypergeometric function.
Then we examine whether or not the
Wronskian vanishes.
However, the existence of a zero frequency quasi-normal mode
does not necessarily imply the onset of instability
because condition (v) does not deny the existence 
of a $\lambda_n^{false}$ for $\omega_n$
such that $\omega_n(\lambda_n^{false})=0$
but $\mbox{Im}[\omega_n(\lambda)]>0$ 
for $\lambda>\lambda_n^{false}$
and also for $\lambda<\lambda_n^{false}$.
In order to check the onset of instability 
we relax the assumption $\omega=0$.
We consider a pure imaginary frequency, i.e.,
\begin{equation}
  \label{pureimaginary}
  \omega= - i c,~~( c>0 )  
\end{equation}
and under the regularity condition
we integrate Eq.~(\ref{eq:schroedinger})
from the origin to the surface and thereafter
from the surface to the asymptotic 
region $|\omega|r\gg1$ and $r\gg M$.
Then we calculate the Wronskian 
with the asymptotic expansion of $\psi^{-}$
and the coefficient $C^{+}(\omega)$.
The details of the asymptotic expansion
are described in Appendix \ref{sec:asymptoticexpansion}.
When $\beta_0$ is changed to a slightly smaller (larger) value than
the value for the emergence of a zero-frequency mode,
we can confirm that
an unstable quasi-normal mode emerges
by observing that
$C^{+}(\omega=- i c)$ crosses zero as $c$ is increased from zero.

\begin{figure}[htb]
  \epsfysize 6cm \epsfxsize 8cm
  \centerline{\epsfbox{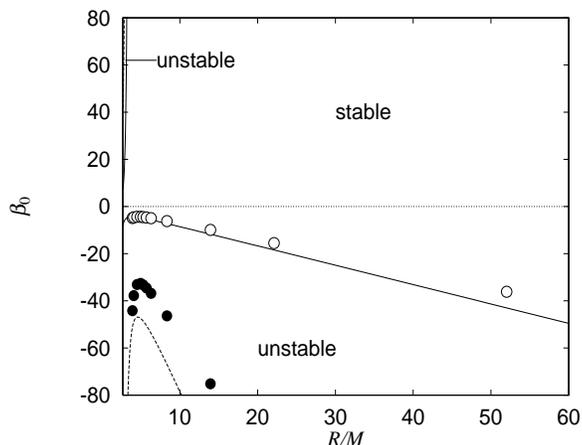}}
\caption{Stability versus instability.
The ordinate is the theoretical parameter $\beta_0$
and the abscissa is the stellar radius.
The solid lines denote the lines $\beta_0=\beta_0^{crit}$,
and the dashed lines denote the onset of instability 
of another quasi-normal mode in the incompressible fluid model.
The open circles denote the points $\beta_0=\beta_0^{crit}$,
and the filled circles the onset of instability
of another quasi-normal mode in the polytropic model.}
\end{figure}

\begin{figure}[htb]
  \epsfysize 6cm \epsfxsize 8cm
  \centerline{\epsfbox{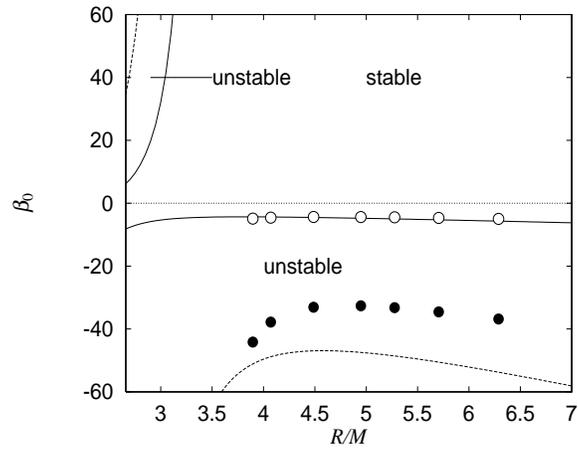}}
  \caption{Magnification of the region of smaller stellar radius
in Fig. 4.}
\end{figure}

\begin{table}[htb]
  \begin{center}
    \caption{The critical values for the polytropic stellar models.
$\rho_c$, $M$ and $R$ are the total energy density at the center,
gravtitational mass and radius, respectively.
Only the equilibrium solutions that satisfy the necessary 
condition for stability
$dM/d\rho_c>0$ are in the table.}
    \begin{tabular}{lccr} \hline \hline
      $\rho_c(\mbox{g/cm}^3)$ & $M/M_{\odot}$ 
      & $R(\mbox{km})$ & $\beta_0^{crit}$
      \\ \hline
      2.26E+15 & 1.95 & 11.2 & -4.90 \\
      1.92E+15 & 1.93 & 11.6 & -4.59 \\
      1.45E+15 & 1.85 & 12.2 & -4.35 \\
      1.16E+15 & 1.74 & 12.7 & -4.39 \\
      1.03E+15 & 1.67 & 12.9 & -4.48 \\
      8.96E+14 & 1.57 & 13.2 & -4.67 \\
      7.70E+14 & 1.45 & 13.4 & -4.97 \\
      5.32E+14 & 1.12 & 13.7 & -6.21 \\
      3.10E+14 & 0.670 & 13.7 & -9.94 \\
      2.04E+14 & 0.410 & 13.3 & -15.5 \\
      1.00E+14 & 0.161 & 12.3 & -36.2 \\ \hline
    \end{tabular}
  \end{center}
\end{table}

The results of numerical calculations are shown in Figs. 4, 5 and 6.
Figures 4 and 5 are diagrams of stability versus instability 
in the $\beta_0$ vs. $R/M$ plane.
The solid and dashed lines denote the emergence
of the first and second unstable modes 
for the incompressible fluid model.
Here we have restricted our attention to the 
stellar model satisfying $R>(8/3)M$
so that the dominant energy condition may be satisfied.
If we consider more realistic models,
there is the necessary condition, $dM/d\rho_c >0$, 
for stability,
where $\rho_c$ is the total density at the center. 
From this condition, 
the ratio $R/M$ cannot 
be smaller than some critical value
that depends on the equation of state,
for example, $3.18 - 4.63$ for models in Ref.~\cite{arnett-bowers}.
If the radius is smaller than this critical radius,
the stellar model 
is already unstable
even if $\beta_0=0$, i.e., even if there is no 
coupling between the scalar field and matter.
The unstable region above the line $\beta_0=0$, 
as seen in Figs. 4 and 5, is due to the
violation of negativity of $T=-\rho+3P$
for the stellar radius $R<(18/5)M$.
As seen in Fig. 4, for $R\gg M$ the curve of 
$\beta_0=\beta_0^{crit}$
can be fitted by the straight line 
\begin{equation}
  \label{eq:fitting}
  \beta_0^{crit}=-\frac{\pi^2}{12}\frac{R}{M},
\end{equation}
and this line is derived by the emergence of 
the unstable mode in the square well potential ignoring
the space-time curvature and matter pressure,
and assuming a homogeneous density distribution.
The open and filled circles denote
the emergence of the first and second 
unstable quasi-normal modes
for the polytropic stellar models that satisfy
the necessary condition for stability, $dM/d\rho_c >0$.
They are also shown in Table I.
Figures 4 and 5 suggest that the 
critical value $\beta_0^{crit}$
for the emergence of the first unstable mode
has little dependence on the
equation of state, as a function of the ratio $R/M$.
Figure 6 displays the emergence of an
unstable quasi-normal mode for the
incompressible fluid model of $R=10M$.
Since the eigenvalue of the
first unstable mode is approximately given
by the depth of the effective potential $V$
and the third term dominates other terms 
in Eq.~(\ref{eq:potential}),
if $\beta_0$ is considerably smaller than $\beta_0^{crit}$, 
the growth time $\tau$
for the first unstable mode is approximately given by
\begin{equation}
  \tau\sim \sqrt{\frac{1}{|\beta_0 T|}}
  \sim \frac{\tau_{\mbox{ff}}}{\sqrt{|\beta_0|}},
\end{equation}
where $\tau_{\mbox{ff}}$ denotes the free-fall time of the star.

\begin{figure}[htb]
  \epsfysize 6cm \epsfxsize 8cm
  \centerline{\epsfbox{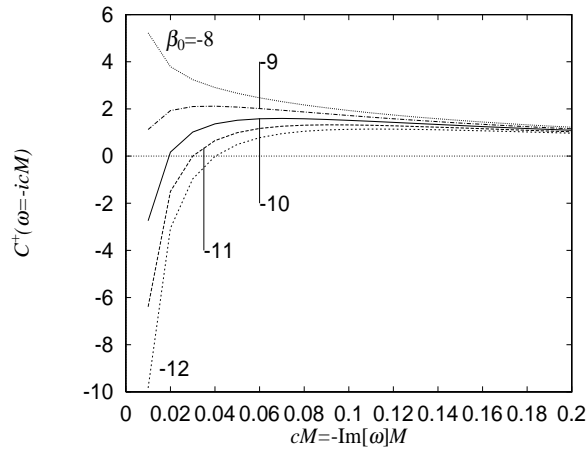}}
\caption{Emergence of an unstable quasi-normal mode.
The ordinate is $C^{+}(\omega=-ic)$ and
the abscissa is $c=-\mbox{Im[}\omega\mbox{]}$.}
\end{figure}
 
\section{SUMMARY AND DISCUSSIONS}
\label{sec:summary}

In the scalar-tensor theories of gravity 
in which $\alpha_0\equiv\alpha(\varphi_0)=0$,
the field equations allow a solution
that consists of a constant scalar field
and a solution of the Einstein 
equation for metric tensor 
and matter.
The field equations for linear perturbations
of this solution
are decoupled to an equation for the
scalar field perturbation and equations for 
the metric and matter perturbations. 
The equations for the metric and matter perturbations
are identical to those in general relativity.

The stability analysis of the equation for the scalar 
field perturbation is given in terms of the 
quasi-normal mode frequencies.
From a general argument on the quasi-normal modes
of the scalar field
in the static and spherically symmetric space-time,
the following results turn out to be true:
An unstable mode grows monotonically.
The presence of a non-negative effective potential 
implies the
absence of unstable quasi-normal modes.
When a quasi-normal mode changes its stability,
its frequency is zero.

From the shape of the effective potential,
we find the following tendencies with 
regard to stability:
As $l$ becomes smaller, 
the quasi-normal modes of the scalar field 
become more likely to be unstable . 
Stability of the quasi-normal modes also depends on the
first derivative $\beta_0$ of the coupling 
function $\alpha(\varphi)$ 
between the scalar field and matter.
As $\beta_0$ becomes smaller, 
the quasi-normal modes of the scalar field become
more likely to be unstable
if $T=-\rho+3P<0$ in the stellar interior. 

In order to seek for the critical value $\beta_0^{crit}$
for the onset of instability of a quasi-normal mode
of the scalar field,
we examined whether or not a zero-frequency 
exists,
because the frequency of the marginally
stable quasi-normal mode that changes its 
stability must be zero.
Then we confirmed the onset of instability by observing the 
emergence of an unstable mode.

Using the incompressible fluid 
model and polytropic model of a star
we found the critical value $\beta_0^{crit}$
numerically as a function of $R/M$.
If $ R/M\simeq 4$ for neutron stars
and if these stellar models approximate such neutron stars well,
our numerical results suggest 
that a constraint on the parameter $\beta_0$
would be $\beta_0 \gsim -5$ from the stability of the
neutron stars.
In the theories in which $\alpha_0=0$ and 
$\beta_0\lsim-5$ (this value depends slightly on the
equation of state for high-density nuclear matter),
the coupling between the scalar field and matter 
may change the stability of a stellar solution
that consists of a solution for the Einstein equation
and a constant scalar field.
Even if a stable equilibrium solution 
exists for $\beta_0 \lsim -5$,
such a solution must contain non-perturbative effects
due to the coupling between the scalar field and 
matter and it will have
significantly different form factors 
for the pulsar-timing experiments
from those in general 
relativity~\cite{damour-esposito-farese3}.
Even for non-relativistic stars, such as white dwarfs,
main-sequence stars, planets and so on,
there is a critical value $\beta_0^{crit}\sim - R/M$
for their stability.

Damour and Esposito-Far\'ese discovered 
that, for a coupling function of the form
\begin{equation}
  \alpha(\varphi)=\beta\varphi,
\end{equation}
when $\beta\lsim -4$, non-perturbative effects develop for
massive neutron stars. 
That is, the star has a 
nontrivial finite ``scalar charge'', and the scalar
field displays nontrivial configuration even when 
the asymptotic value $\varphi_0$
is extremely small or 
zero~\cite{damour-esposito-farese2,damour-esposito-farese3}.
The solution with the non-perturbative effects is significantly different 
from that in general relativity.
These results raise a paradoxical problem 
in the limit $\varphi_0 \to 0$.
If $\varphi_0=0$, both the field equations and boundary condition
possess symmetry under the reflection transformation $\varphi-\varphi_0
\to -(\varphi-\varphi_0)$, and hence they have the trivial solution
$\varphi(x)=\varphi_0=0$, and the matter and metric tensor
satisfy the Einstein equation in general relativity.
It seems that this fact implies a discontinuity in the sequence of the
solutions against $\varphi_0$.
Recently it has been discussed
that the emergence of the non-perturbative effects on
the equilibrium solutions of 
neutron stars is interpreted as a ``spontaneous
scalarization'' analogous to
a spontaneous magnetization of the 
ferromagnets~\cite{damour-esposito-farese3}.
The critical value $\beta_0^{crit} \sim -5 $ for 
the onset of instability
of a quasi-normal mode for relativistic stars
obtained here by our analysis 
is approximately equal to
the critical value in which an equilibrium solution
with non-perturbative effects for relativistic stars 
emerges for 
$\alpha_0=0$~\cite{damour-esposito-farese2,damour-esposito-farese3}.
In the context of spontaneous scalarization,
we have shown that if $\beta_0>\beta_0^{crit}$,
the equilibrium solution with symmetry for the reflection
transformation $\varphi-\varphi_0 \to - (\varphi-\varphi_0)$ is stable,
but if $\beta_0<\beta_0^{crit}$ the solution
with the symmetry is no longer stable.
This fact strongly suggests that a spontaneous
symmetry breaking occurs at $\beta_0=\beta_0^{crit}$.
It is still important to examine the stability of 
the equilibrium solution with non-perturbative effects.

It was shown in Ref.~\cite{harada} that in the 
Oppenheimer-Snyder collapse
the scalar field within the dust
decays with oscillation for 
non-negative values of $\beta_0$
(see Fig. 14 of Ref.~\cite{harada})
and 
grows without oscillation
for a large
negative value of $\beta_0$
(see Fig. 15 of Ref.~\cite{harada}).
This leads to the $\beta_0$-dependence 
of the observed wave form of the 
scalar gravitational waves,
as seen in Figs. 10, 11 and 12 of Ref.~\cite{harada}.
The results of 
the stability analysis given above account for 
the behavior of the scalar field 
in terms of the quasi-normal mode frequencies.
For $\beta_0 > \beta_0^{crit}$ no unstable 
quasi-normal mode exists,
and only the damped oscillation modes are allowed,
but for $\beta_0<\beta_0^{crit}$ 
unstable modes exist and
grow exponentially without oscillation.
Those purely outgoing wave modes 
characterize the time development of
the scalar field within the dust and 
therefore the observed wave form of the scalar 
gravitational waves.
For $\beta_0>\beta_0^{crit}$, the quasi-normal mode
oscillations appear in the observed wave form before
the last quasi-normal mode ringing of the formed
black hole, as seen in Fig. 11 of Ref.~\cite{harada}.
For $\beta_0<\beta_0^{crit}$, the growing quasi-normal
mode appears in the wave form before the last
quasi-normal mode ringing of the black hole,
as seen in Fig. 12 of Ref.~\cite{harada}.  

Although we have assumed implicitly that 
the unperturbed solution has a regular origin,
we can extend the results obtained in 
Sec. \ref{sec:stabilityanalysis} to a
solution that has an event horizon.
In this case we have only to 
replace the regularity condition 
at the origin $r=0$ with the ingoing wave condition
at the horizon, $r_* \to -\infty$.
If we consider a Schwarzschild or Reissner-Nordtstr\"om 
black hole,
coupling between the scalar fileld and matter
does not exist because $T=0$ for the
vacuum and electromagnetic field.
Therefore the effective potential (\ref{eq:potential})
is identical to that for a non-gravitational scalar field 
in general relativity.
There are, however, some exotic black hole solutions
in which $T\ne0$, and the stability of such black holes
may be altered due to the coupling between 
the gravitational scalar field and matter. 

Although we have assumed $\alpha_0=0$,
the stability analysis given here applies to the
perturbation equations truncated up to zeroth order
in $\alpha_0$ expansions around general 
relativity~\cite{harada}, if $\alpha_0 \neq 0$.
When we need to consider the effect of the nonzero
value of $\alpha_0$ seriously,
we must treat the coupled equations for the perturbations
of the scalar field, metric tensor and matter fields. 
The assumption that
$\alpha_0$ is
extremely small seems to be reasonable 
if the first derivative $\beta_0$ 
of the coupling function $\alpha(\varphi)$ is positive or if
the scalar field has a small mass $m_{\varphi}$
( $10^6\mbox{km} \lsim \hbar /m_{\varphi} \lsim H_0^{-1}$ )
and the location $\varphi_m$ of the minimum of the
potential of $\varphi$ , $V(\varphi)=
(1/8\pi)m_{\varphi}^2(\varphi-\varphi_m)^2$,
coincides with a zero of the coupling 
function $\alpha(\varphi)$,
because in these 
cases the cosmological attraction 
mechanism is effective and $\alpha_0$
is attracted toward 
zero~\cite{damour-esposito-farese3,damour-nordtvedt}.

\acknowledgements

I would like to thank H. Sato, T. Nakamura, K. Nakao,
M. Siino, T. Tanaka and T. Chiba for helpful discussions.
I am also grateful to H. Sato for his 
continuous encouragement.
 
\appendix

\section{EXPANSION OF REGULAR SOLUTION AROUND THE ORIGIN}
\label{sec:regularity}
To suppress numerical errors 
in the stellar interior we integrate not $\psi$
but $u$ defined as
\begin{equation}
  \label{defineu}
  \psi  \equiv r^{l+1}u. 
\end{equation}
Then the differential equation for $u$ is
\begin{eqnarray}
u^{\prime\prime}&+&\left[(\Phi^{\prime}-\Psi^{\prime})
+\frac{2(l+1)}{r}\right]u^{\prime} \nonumber \\
&+&\left[\omega^2 e^{2\Psi-2\Phi}+\frac{\Phi^{\prime}-\Psi^{\prime}}{r}l
-\frac{l(l+1)}{r^2}(e^{2\Psi}-1)+4\pi\beta_0 T e^{2\Psi} \right]u=0.
\end{eqnarray}
In the exterior vacuum region the 
differential equation for $\psi$ is
\begin{eqnarray}
  \label{eq:odeexterior}
    \psi^{\prime\prime}&+&\left(1-\frac{2M}{r}\right)^{-1}\frac{2M}{r^2}
    \psi^{\prime} \nonumber \\
    &+&\left(1-\frac{2M}{r}\right)^{-2}\left[\omega^2-\left(\frac{2M}{r^3}
        +\frac{l(l+1)}{r^2}\right)\left(1-\frac{2M}{r}\right)\right]\psi=0.
\end{eqnarray}

In order to obtain the expansion
of the solution $u$ around the origin,
we expand $T$, $\Phi$ and $\Psi$ around the origin as
\begin{eqnarray}
\label{eq:expt}
T&=&T_0+T_2 r^2 +T_4 r^4 + \cdots ,\\
\label{eq:expphi}
\Phi&=&\Phi_0+\Phi_2 r^2 +\Phi_4 r^4 +\cdots ,\\
\label{eq:exppsi}
\Psi &=& \Psi_0 + \Psi_2 r^2 +\Psi_4 r^4+\cdots.
\end{eqnarray}
If we take the incompressible fluid model,
the coefficients are given by
\begin{eqnarray}
T_0&=&-2\frac{3\gamma-2}{3\gamma-1}\rho_0, \\
T_2&=&-8\pi\frac{\gamma}{(3\gamma-1)^2}\rho_0^2, \\
\Phi_0&=&\ln \frac{3\gamma-1}{2}, \\
\Phi_2&=&\frac{4\pi}{3(3\gamma-1)}\rho_0, \\
\Phi_4&=&\frac{8}{9}\pi^2\frac{3\gamma-2}{(3\gamma-1)^2}\rho_0^2, \\
\Psi_0&=&0, \\
\Psi_2&=&\frac{4}{3}\pi\rho_0, \\
\Psi_4&=&\frac{16}{9}\pi^2\rho_0^2,
\end{eqnarray}
where we have defined
\begin{equation}
  \label{gamma}
  \gamma \equiv \sqrt{1-\frac{2M}{R}},
\end{equation}
and $1/3<\gamma<1$, as seen from Eq.~(\ref{eq:lowestradius}).

Using these expansions, the regular solution
$u$ is expanded around the origin as
\begin{equation}
  \label{expansion}
   u=\sum_{\nu=0}^{\infty}a_\nu r^\nu, 
\end{equation}
where the coefficients $\{a_{\nu}\}$ are determined 
using the coefficients in 
Eqs. (\ref{eq:expt}), (\ref{eq:expphi}) and (\ref{eq:exppsi}) as
\begin{eqnarray}
a_1 &=& a_3=\cdots=0 ,\\
a_2 &=& -\frac{1}{2(2l+3)}\left[
\omega^2 e^{-2\Phi_0}+2l\Phi_2-2l(l+2)
\Psi_2+4\pi\beta_0T_0\right]a_0 ,\\
a_4 &=& -\frac{1}{4(2l+5)}\left\{\left[
2\omega^2e^{-2\Phi_0}(\Psi_2-\Phi_2)+4 l(\Phi_4-\Psi_4)\right.\right.
\nonumber \\
& & \left.\left.
-2l(l+1)(\Psi_4+\Psi_2^2)+4\pi\beta_0(T_2+2T_0\Psi_2)\right]a_0\right.
\nonumber \\
& & \left.
+\left[\omega^2 e^{-2\Phi_0}+2(l+2)\Phi_2
-2(l^2+2l+2)\Psi_2
+4\pi\beta_0T_0\right]a_2\right\} \\
&=& \frac{1}{8(2l+3)(2l+5)}\left\{\left[
\omega^2 e^{-2\Phi_0}+(4+2l)\Phi_2
-2(l^2+2l+2)\Psi_2\right.\right.
\nonumber \\
& & \left.\left.+4\pi\beta_0T_0\right]
\left[\omega^2e^{-2\Phi_0}+2l\Phi_2
-2l(l+2)\Psi_2+4\pi\beta_0T_0\right]\right.
\nonumber \\
& & \left.
-2(2l+3)\left[2\omega^2 e^{-2\Phi_0}
(\Psi_2-\Phi_2)+4l(\Phi_4-\Psi_4)\right.\right.
\nonumber \\
& & \left.\left.
-2l(l+1)(\Psi_4+\Psi_2^2)+4\pi\beta_0
(T_2+2T_0\Psi_2)\right]\right\}a_0 ,\\
a_6 &=& \cdots, \nonumber 
\end{eqnarray}
and so on. 
  
\section{NUMERICAL CALCULATION OF THE POLYTROPIC STELLAR MODEL}
\label{sec:poly}
We make the variables dimensionless:
\begin{eqnarray}
  \tilde{n} &\equiv& \frac{n}{n_0} = \frac{n}{0.1\mbox{fm}^{-3}}, \\
  \tilde{\rho} &\equiv& \frac{\rho}{n_0m_b} 
  = \frac{\rho}{1.66\times10^{14}\mbox{gcm}^{-3}}, \\
  \tilde{P} &\equiv& \frac{P}{n_0m_b}
  = \frac{P}{1.49\times10^{35}\mbox{gcm}^{-1}s^{-2}}, \\
  \tilde{r} &\equiv& \frac{r}{(n_0m_b)^{-1/2}}
  = \frac{r}{9.02\times10^6\mbox{cm}}, \\
  \tilde{m} &\equiv& \frac{m}{(n_0m_b)^{-1/2}}
  = \frac{m}{1.22\times10^{35}\mbox{g}}.
\end{eqnarray}
Hereafter we omit tildes for simplicity.
Then we obtain the following ordinary differential equations for $n$ ,
$m$ and $\Phi$ from Eqs. (\ref{eq:tov}), (\ref{eq:qlmass})
and (\ref{eq:phi}):
\begin{eqnarray}
  \frac{dn}{dr} &=& -\frac{n(P+\rho)}{\Gamma P}
  \frac{m+4\pi r^3 P}{r(r-2m)}, 
  \label{eq:dndr} \\
  \frac{dm}{dr} &=& 4\pi \rho r^2, 
  \label{eq:dmdr} \\
  \frac{d\Phi}{dr} &=& \frac{m+4\pi r^3 P}{r(r-2m)},
  \label{eq:dphidr}
\end{eqnarray}
where 
\begin{eqnarray}
  \rho &=& n+\frac{P}{\Gamma-1}, \\
  P &=& Kn^{\Gamma}.
\end{eqnarray}
We integrate Eqs. (\ref{eq:dndr}), (\ref{eq:dmdr}) and (\ref{eq:dphidr})
from $r=0$. 
The initial values of $n$, $m$ and $\Phi$ are
\begin{eqnarray}
  n &=& n_c, \\
  m &=& 0, \\
  \Phi &=& 0.
\end{eqnarray}
In order to match the internal solution 
to the exterior Schwarzschild
space-time at the stellar surface,
we modify the calculated value $\Phi_{calc}$ to 
the true value of $\Phi_{true}$ by
\begin{equation}
  \Phi_{true}(r)=\Phi_{calc}(r)-\Phi(R)_{calc}
  +\frac{1}{2}\ln\left(1-\frac{2M}{R}\right).
\end{equation}
To guarantee regularity at the origin,
we use the following
expansions of the solution around the origin:
\begin{eqnarray}
  n &=& n_c+n_2 r^2+ n_4 r^4+\cdots, \\
  m &=& m_3 r^3+ m_5 r^5+ \cdots, \\
  \Phi &=& \Phi_2 r^2 +\Phi_4 r^4 +\cdots, \\
  \rho &=& \rho_c + \rho_2 r^2 +\cdots, \\
  P &=& P_c +P_2 r^2 +P_4 r^4 +\cdots. 
\end{eqnarray}
Then the coefficients are given by the following
relations.
\begin{eqnarray}
  P_c &=& Kn_c^{\Gamma}, \\
  \rho_c &=& n_c + \frac{P_c}{\Gamma-1}, \\
  m_3 &=& \frac{4}{3}\pi\rho_c, \\
  n_2 &=& -n_c\frac{P_c+\rho_c}{2\Gamma P_c}(m_3+4\pi P_c), \\
  P_2 &=& \Gamma P_c \frac{n_2}{n_c}, \\
  \rho_2 &=& n_2 + \frac{P_2}{\Gamma-1}, \\
  m_5 &=& \frac{4}{5}\pi \rho_2, \\
  \Phi_2 &=& -\frac{P_2}{P_c+\rho_c}, \\
  n_4 &=& -\frac{1}{4\Gamma P_c}\left[\left\{
      n_2(P_c+\rho_c)+n_c\left(\rho_2-\frac{\rho_c}{P_c}P_2\right)
      \right\}(m_3+4\pi P_c)\right.\nonumber \\
      & & \left.+n_c(P_c+\rho_c)(m_5+4\pi P_2)
      +2n_c(P_c+\rho_c)(m_3+4\pi P_c)m_3\right],\\
  P_4 &=& \frac{1}{2}\Gamma(\Gamma-1)P_c\left(\frac{n_2}{n_c}
    \right)^2+\Gamma P_c\left(\frac{n_4}{n_c}\right),\\
  \Phi_4 &=& -\frac{P_4}{P_c+\rho_c}
  +\frac{P_2(P_2+\rho_2)}{2(P_c+\rho_c)^2}.
\end{eqnarray}
\section{STATIC SOLUTION IN THE EXTERIOR SCHWARZSCHILD SOLUTION}
\label{sec:exterior}
If we assume the scalar field $\psi$ does not depend on $t$, 
Eq.~(\ref{eq:schroedinger}) reduces to
\begin{equation}
  \label{eq:omega0}
\frac{d^2\psi}{dr^2}+\left(1-\frac{2M}{r}\right)^{-1}\frac{2M}{r^2}
\frac{d\psi}{dr}-\left(\frac{2M}{r^3}+\frac{l(l+1)}{r^2}\right)
\left(1-\frac{2M}{r}\right)^{-1}\psi=0.
\end{equation}
Here we define
\begin{equation}
  \label{eq:zeta}
  \zeta\equiv\frac{2M}{r},
\end{equation}
and
\begin{equation}
  \label{eq:w}
  w\equiv\frac{\psi}{\zeta^{l}}.
\end{equation}
Then Eq.~(\ref{eq:omega0}) transforms into
\begin{equation}
  \label{eq:hypergeometriceq}
  \zeta(1-\zeta)\frac{d^2w}{d\zeta^2}+[(2+2l)-(2l+3)\zeta]
  \frac{dw}{d\zeta}-(l+1)^2 w=0.
\end{equation}
This is the hypergeometric equation. 
Regularity at $r=\infty$ requires
\begin{equation}
  \psi\propto \frac{1}{r^l},
\end{equation}
in the limit $r\to \infty$.
When we require this condition, the unique
solution can be written as
\begin{equation}
  \psi=q \left(\frac{2M}{r}\right)^l
  F\left(l+1,l+1,2l+2;\frac{2M}{r}\right),
\end{equation}
where $F$ is the hypergeometric function
and $q$ is an arbitrary constant.
\section{ASYMPTOTIC EXPANSIONS}
\label{sec:asymptoticexpansion}
In Eq.~(\ref{eq:odeexterior}), if $\omega\neq0$,
$r=\infty$ is an irregular singular point.
A general solution is expressed as a 
linear combination of the purely ingoing 
and outgoing waves, as seen in Eq.~(\ref{eq:linearcomb}).
The asymptotic expansions
for $\psi^{+}$ and $\psi^{-}$
in the asymptotic region $|\omega|r\gg 1$ 
and $r\gg M$ are  
  \begin{eqnarray}
    \psi^{+}(r_*) &=& e^{i\omega
      r_*}\sum_{\nu=0}^{\infty}b_{\nu}^{+}r^{-\nu},\\
    \psi^{-}(r_*) &=& e^{-i\omega
      r_*}\sum_{\nu=0}^{\infty}b_{\nu}^{-}r^{-\nu},
  \end{eqnarray}
where the coefficients $\{b_{\nu}^{\pm}\}$ are given by
the following recursive relations:
  \begin{eqnarray}
    b_1^{\pm} &=& \pm i \frac{l(l+1)}{2\omega}b_0^{\pm},\\
    b_2^{\pm} &=& \frac{1}{8\omega^2}[-(l-1)l(l+1)(l+2)
    \pm i 4\omega M ]b_0^{\pm},\\
    b_{\nu}^{\pm}&=&\mp\frac{i}{2\omega\nu}\left[
      \{(\nu-1)\nu-l(l+1)\pm i 4\omega M(\nu-1)\}b_{\nu-1}
      ^{\pm}\right. \nonumber \\
      & & -2M\{(\nu-1)(2\nu-3)-l(l+1)\}b_{\nu-2}^{\pm}
        \\ \nonumber
      & & \left.+4M^{2}(\nu-2)^2b_{\nu-3}^{\pm}\right]
      ~~~~~~\mbox{for}~\nu\geq 3.
  \end{eqnarray}

\end{document}